\documentclass[twocolumn,prb,amsmath,amssymb,amsfonts,superscriptaddress,floatfix,nopacs]{revtex4}

\usepackage{graphicx}
\usepackage{dcolumn}
\usepackage{bm}
\usepackage{epstopdf}
\usepackage{color}

\def\S{\rm S}
\def\be{\begin{equation}}
\def\ee{\end{equation}}
\def\bd{\begin{displaymath}}
\def\ed{\end{displaymath}}
\def\-{\phantom{-}}

\begin{document}

\title{Spatial dependence of the superexchange interactions for transition-metal trimers in graphene}

\author{Charles B. Crook}
\affiliation{Department of Physics and Astronomy, James Madison University, Harrisonburg, VA 22802}
\author{Gregory Houchins}
\affiliation{Department of Physics and Astronomy, James Madison University, Harrisonburg, VA 22802}
\author{Jian-Xin Zhu}
\affiliation{Theoretical Division, Los Alamos National Laboratory, Los Alamos, NM 87545, USA}
\affiliation{Center for Integrated Nanotechnologies, Los Alamos National Laboratory, Los Alamos, NM 87545, USA}
\author{Alexander V. Balatsky}
\affiliation{Institute for Materials Science, Los Alamos National Laboratory, Los Alamos, NM 87545, USA}
\affiliation{NORDITA, Roslagstullsbacken 23, 106 91 Stockholm, Sweden}
\author{Costel Constantin}
\affiliation{Department of Physics and Astronomy, James Madison University, Harrisonburg, VA 22802}
\author{Jason T. Haraldsen}
\affiliation{Department of Physics, University of North Florida, Jacksonville, FL 32224}

\date{\today}

\begin{abstract}
This study examines the magnetic interactions between spatially-variable manganese and chromium trimers substituted into a graphene superlattice. Using density functional theory, we calculate the electronic band structure and magnetic populations for the determination of the electronic and magnetic properties of the system. To explore the super-exchange coupling between the transition-metal atoms, we establish the magnetic magnetic ground states through a comparison of multiple magnetic and spatial configurations. Through an analysis of the electronic and magnetic properties, we conclude that the presence of transition-metal atoms can induce a distinct magnetic moment in the surrounding carbon atoms as well as produce an RKKY-like super-exchange coupling. It hoped that these simulations can lead to the realization of spintronic applications in graphene through electronic control of the magnetic clusters. 
\end{abstract}

\pacs{Valid PACS appear here}
\keywords{Graphene, Superexchange, RKKY Interaction, Density Functional Theory}
\maketitle

\section{Introduction}

Two-dimensional (2D) materials are producing a revolution in materials engineering and design that is hoped to have a large influence on the advancement of technology through enhancements and understanding of electron interactions\cite{fio:14,wolf:01,awsc:02,awsc:07,li:16,wehl:14}. Recently, there has been large push to incorporate magnetic and spin degrees of freedom to make spintronic or magnonic materials\cite{bade:10,zuti:04,khaj:11,xu:06,sato:02}, which is an area being explored for its potential applications in quantum computing as well as memory and storage\cite{enge:01,niel:10}. 

Graphene is the most well-known 2D Dirac material and has gained a significant amount of attention due to its high tensile strength and large electron mobility.  The electron interactions in graphene are mediated through the carbon $\pi$-bonds produced in its honeycomb lattice\cite{boehm:62,geim:07,neto:09,pop:12,bolo:08,moro:08}. These properties and its ability to conduct electrons and heat has made graphene a strong candidate for electronic devices\cite{lee:08,sava:12}. However, to use graphene in the development of any spintronic device, one must be able to make it magnetic or couple it to magnetic spin\cite{yazy:08,rao:12,pesi:12,sher:11,duga:06}, which has been achieved in many ways. These include placing a graphene sheet on YIG (yttrium iron garnet)\cite{wang:15}, addition of hydrogen on the surface\cite{gon:16,bou:08}, as well as the substitution of a magnetic impurities\cite{crook:15,hu:11,sant:10,tha:16}.

\begin{figure}
\includegraphics[width=1.0 \linewidth]{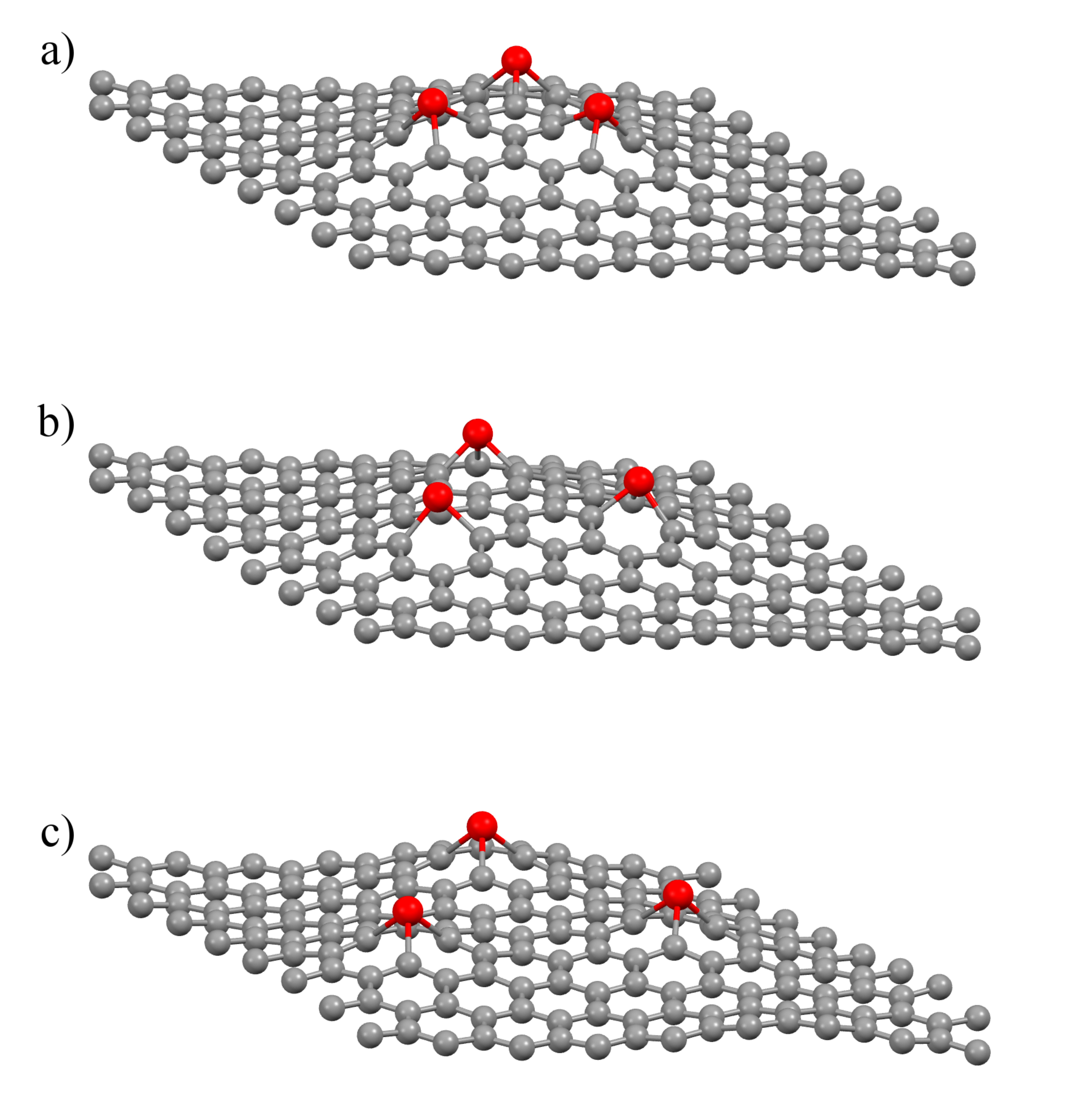}
\caption{Three dimensional view of the distorted graphene structure with the magnetic trimer separated by 3 carbons (a), 5 carbons (b), and 7 carbons (c).}
\label{tilt}
\end{figure}

In recent years, there have been many theoretical and computational studies that have shown that it is possible to substitute the carbon atoms in graphene with magnetic ones\cite{crook:15,hu:11,sant:10,tha:16,saha:10}. Furthermore, an experimental study by Gomes $et~al.$ showed that direct substitution using scanning tunneling microscopy methods was also possible\cite{gom:12}. Therefore, it is attractive and practical to examine the interactions of magnetic atoms that are directly substituted into the graphene lattice. This will maximize orbital overlap and provide a better understanding of spin-spin exchange interaction through the carbon bonds. In a previous article, the interactions of a magnetic dimer were considered. Therefore, it is important to consider the systematic effects of three magnetic spins (a magnetic trimer) to investigate multiple interactions on the ground state\cite{crook:15}.

\begin{figure}
\includegraphics[width=1.00 \linewidth]{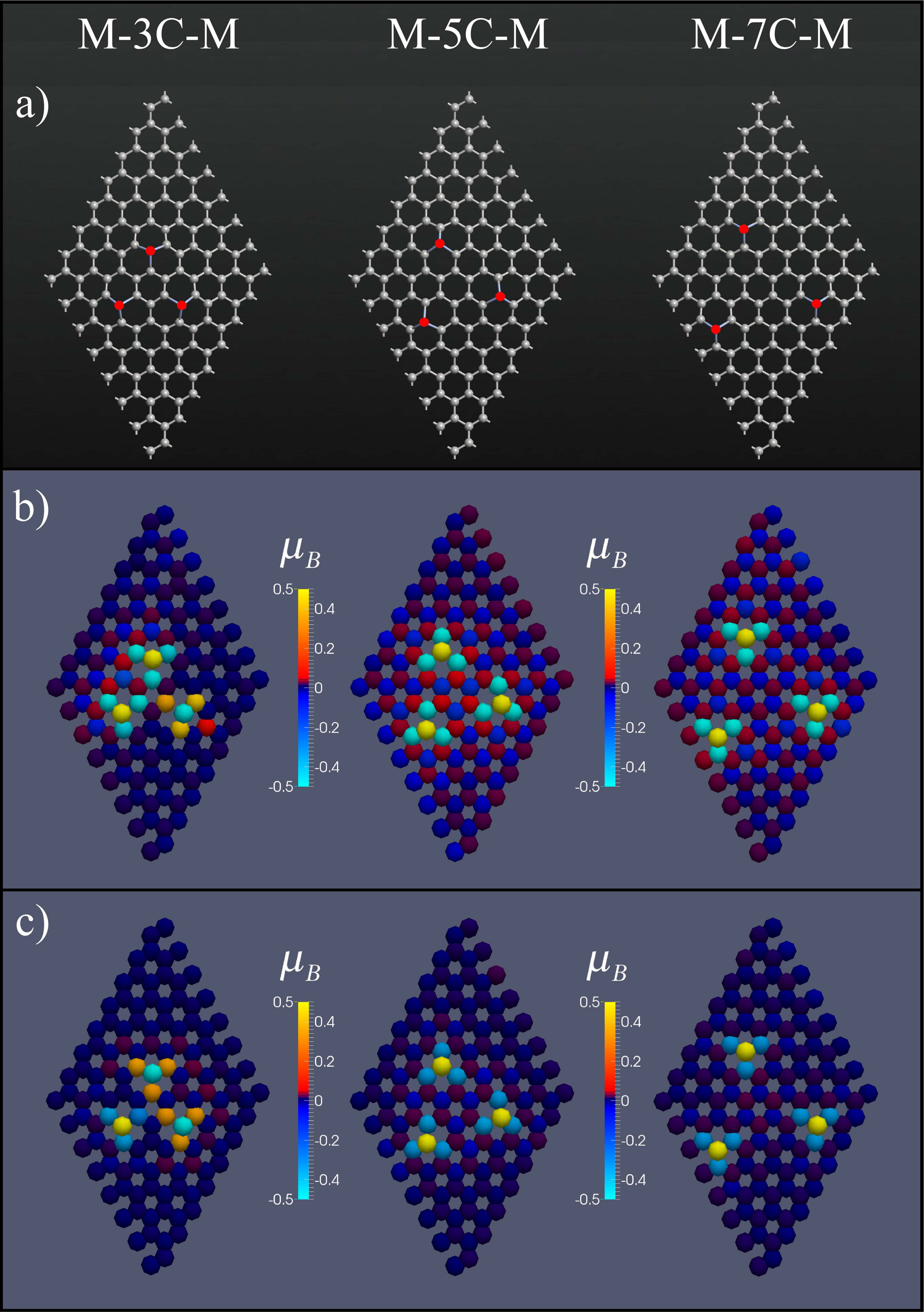}
\caption{a) The general substitution positions for the three different spatial separations, 3, 5, and 7 carbon atoms. The magnetic moment mapped onto the transition-metal substituted graphene lattice with separations between 3 Cr atoms (b) and 3 Mn atoms (c) for each of the three separations. Since they are off the scale, the magnetic moments of the Cr and Mn are 4.2 and 4.0 $\mu_B$, respectively.}
\label{mag}
\end{figure}

In this article, we examine the effects of spatial distribution on transition-metal trimers on the electronic and magnetic properties of graphene in the hopes of producing an electronically controllable magnetic state. Using a large graphene supercell, we introduce three magnetic impurities into the graphene lattice in an equilateral configuration with variable spatial configurations (shown in Fig. \ref{tilt}). Through a comparison of the total energy of different magnetic configurations, we are able to determine the magnetic ground state and estimate the exchange interaction. Here, it is determined that the exchange interactions varies with spatial distance. Furthermore, from the magnetic population, it is shown that the carbon atoms has an induced magnetic moment, where an analysis of the electronic structure and density of states indicates the exchange coupling is related to the conduction electrons in the carbon atoms. This leads to the conclusion that there may exist an RKKY-like super-exchange interaction between the magnetic atoms, which has the potential for control and tunability of the magnetic states with external fields.

\section{Computational Methods}

To examine the electronic and magnetic states, we performed density functional theory (DFT) calculations using Atomistix Toolkit\cite{quantumwise,bran:02,sole:02}. These calculations used a spin-polarized generalized gradient approximation along with an onsite Hubbard-U (SGGA+U) in a LCAO basis using PBE (Perdew, Burke, and Ernzerhof) functionals\cite{perd:96}. The 2D honeycomb lattice of a graphene supercell with three transition-metal atoms substituted in such a way that N carbon atoms separate them through minimum distance, where N = 3, 5, and 7 (Fig. \ref{tilt}). 

Using a full geometry optimization within a 3x3x1 $k$-point sampling, the DFT calculations used a self-consistent energy tolerance of 0.30 $\mu$eV with an energy cutoff of 2.0 keV. {Recent calculations on a magnetically-substituted graphene nanoribbons showed that the implementation of a Hubbard U provides a clearer magnetic ground state when used in a device configuration\cite{houc:17}. Therefore, while the electronic difference between a U = 0 and U = 4 eV is negligible, a Hubbard onsite potential of 4.0 eV was applied to the $d$-orbitals of each magnetic atom, since transition-metal typically need a finite level of potential in comparison to experiment.} 

\begin{figure}
\includegraphics[width=1.0 \linewidth]{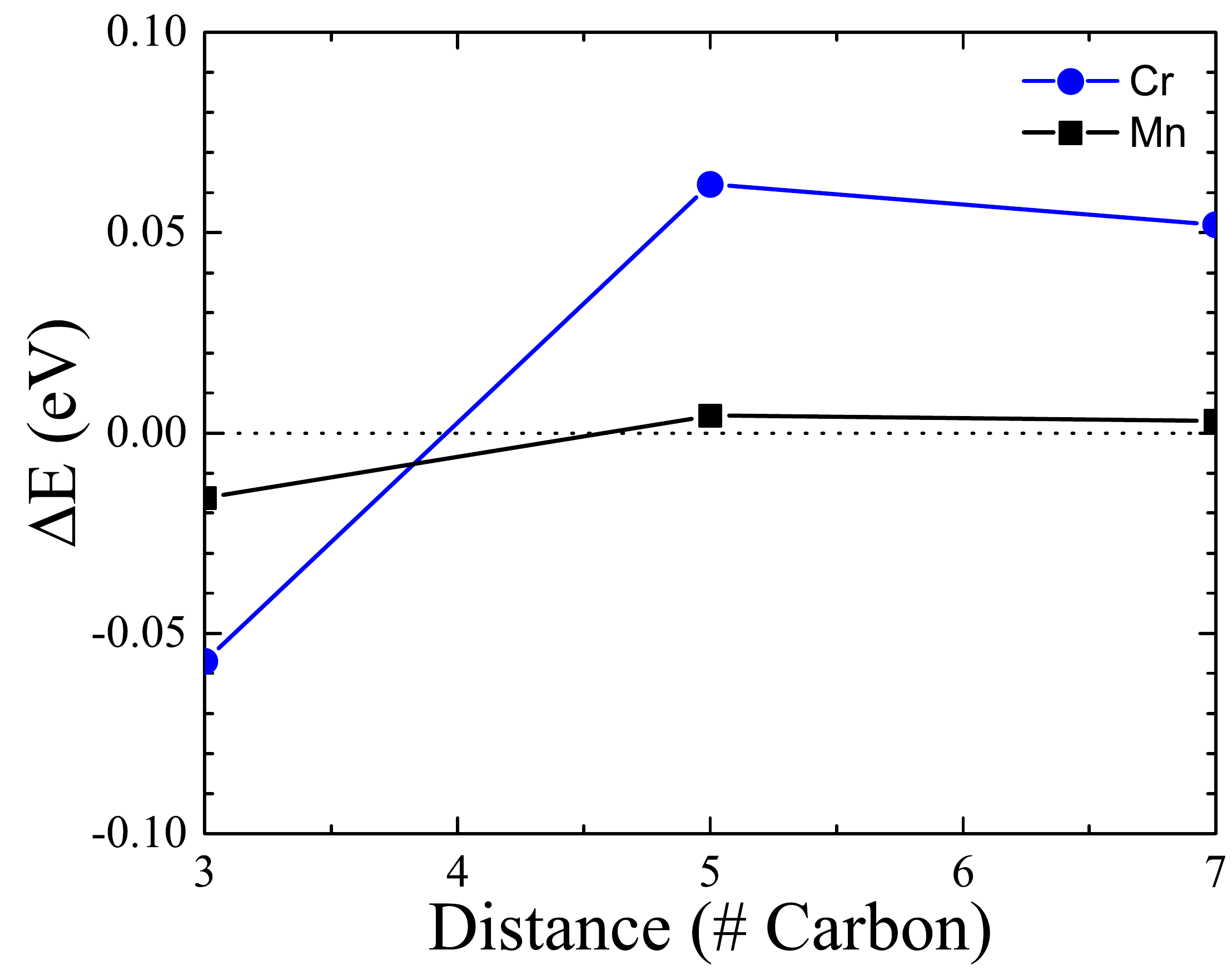}
\caption{Calculated {change in energy} for Cr (blue circles) and Mn (black squares). For each magnetic atom species, there is a change from an anti-ferromagnetic (AFM) to a ferromagnetic (FM) state.}
\label{ground}
\end{figure}

The initial magnetic configurations of up-up-up ($\uparrow \uparrow \uparrow$), up-up-down ($\uparrow \uparrow \downarrow$), and up-down-down ($\uparrow \downarrow \downarrow$) were placed on the transition-metal trimers, while carbon atom were set to an initial spin-zero state. The magnetic configurations of $\uparrow \uparrow \uparrow$ and $\downarrow \downarrow \downarrow$ were found to be equivalent, which is expected given the symmetry of the system. In all, we considered eighteen total configurations (three spatial configurations, three magnetic configurations, two different types of magnetic atoms). 

The magnetic ground states for each configuration were determined through a comparison of the total energies for each configuration. The difference between the antiferromagnetic and ferromagnetic configurations was used to estimate the exchange parameters between the three magnetic atoms. Through an analysis of the electronic band structure and density of states (DOS), we are able to assess the magnetic effect on the Dirac cone at the $K$ point in the Brillouin zone as well as the affects on the electronic transport within the 2D lattice.\cite{crook:15}

\section{Magnetization mapping}

To examine the overall magnetic moment on the transition-metal atoms, Figure \ref{mag} shows magnetization mapping of the Mulliken population for each atom in the graphene supercell, where the transition-metal atoms are depicted by the extreme ends of the scale. Two different magnetic atoms were used to make the trimer cases, Cr and Mn (Figure \ref{mag}(b) and (c)). In the cases for Cr, the interaction between the Cr atoms switch from an AFM to a FM character as the trimer is enlarged. The same type of interaction switch occurs in the Mn case. The calculated total energy for the ground states and the switch can be seen in Figure \ref{ground}. The color legend next to these mappings denote the magnetization within the carbon atoms, and the scale does not give the magnetization of the magnetic atoms. The magnetic moments of the Cr and Mn are 4.2 and 4.0 $\mu_B$ respectively. 

The change in magnetic character (antiferromagnetism to ferromagnetism) coupled to the presence of conduction electrons seems to indicate the possibility of an RKKY interaction. Although, it is clearly much weaker for the Mn case, which may be due to the localization of electrons.


\section{Effective Spin-Spin Exchange Modeling}

To understand the magnetic ground states for the isolated spin clusters embedded in graphene, we determined the total energy for multiple configurations and compared them to the excitation spectrum using Heisenberg spin-spin exchange model. {The spin Hamiltonian can be described as
\be
{\cal H}
=
-\frac{J}{2}\,
\Big(
\vec{\S}_1 \cdot \vec{\S}_2 + \vec{\S}_2 \cdot \vec{\S}_3 +
\vec{\S}_3 \cdot \vec{\S}_1
\Big)\ ,
\label{EtrimerH}
\ee
where the super-exchange constant $J$ is} negative for antiferromagnetic interactions and positive for ferromagnetic ones, and $\vec {\rm S}_i$ is the quantum spin operator\cite{hara:05}.{Within the classical limit, the energy for collinear states can be described by
\be
E = -\frac{JS^2}{2} \big(\cos(\Delta\theta_{12})+\cos(\Delta\theta_{23})+\cos(\Delta\theta_{13}))
\ee
where the ferromagnetic case ($\uparrow \uparrow \uparrow$) is $-\frac{3JS^2}{2}$ and the antiferromagnetic case ($\uparrow \uparrow \downarrow$) is $-\frac{JS^2}{2}$. By examining the difference in energy ($\Delta E$) between the $\uparrow \uparrow \uparrow$ and $\uparrow \uparrow \downarrow$ configurations for classical model is proportional to $J$\cite{hara:09,hara:16}. Similarly, the quantum triangle also produces an energy gap proportional $J$. Therefore, by examining the change in the total energy of the two configurations from density functional theory, we can estimate the exchange interaction between the spin moments as well as investigate the change in this energy as the separation between transition metal atoms is increased, which will provide us with a spatial response of the exchange interaction.}


Figure \ref{ground} shows the calculated change in energy as a function of spatial distance regarding carbon atoms between the magnetic atoms. It is noted that the Cr trimers produce a larger magnetic exchange interaction than the Mn trimers. Furthermore, there is a change from antiferromagnetic to ferromagnetic as the atomic distance is increased. It is important to note that for the largest separation of the 7 carbon atoms the periodicity of the lattice produces linear chain effect. This explains why the 7 atom separation has a higher change in energy and still favors the ferromagnetic configuration. Either way, this response is similar to an RKKY interaction, which is also simulated in dimers\cite{crook:15}. However, to be a true RKKY interaction, the super-exchange interactions must be mediated through the conduction electrons.

\begin{figure}
\includegraphics[width=1.1 \linewidth]{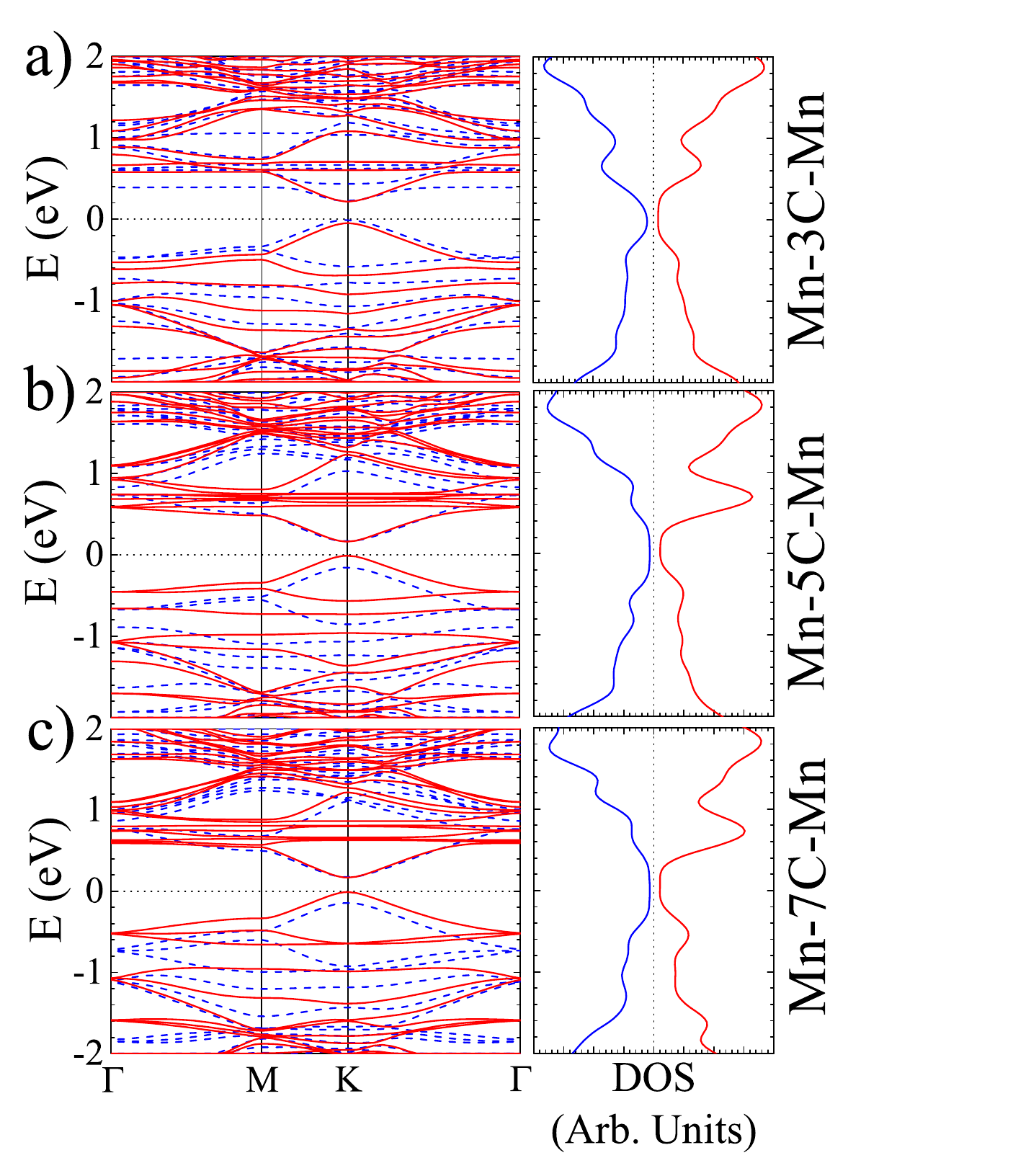}
\caption{The electronic band structure and density of states in manganese-substituted graphene with 3 (a), 5 (b), and 7 (c) separations.}
\label{Mn}
\end{figure}

\section{Electronic structure and density of states}

Figure \ref{Mn} shows the electronic band structure and DOS of the three spatial configurations for the Mn trimers. The DOS shows the presence of a small energy gap at the Fermi level, which means the system is not metal and is more semiconducting. Therefore, the electrons do not have the conduction ability to utilize the RKKY interaction and the exchange energy drops dramatically. 

\begin{figure}
\includegraphics[width=1.1 \linewidth]{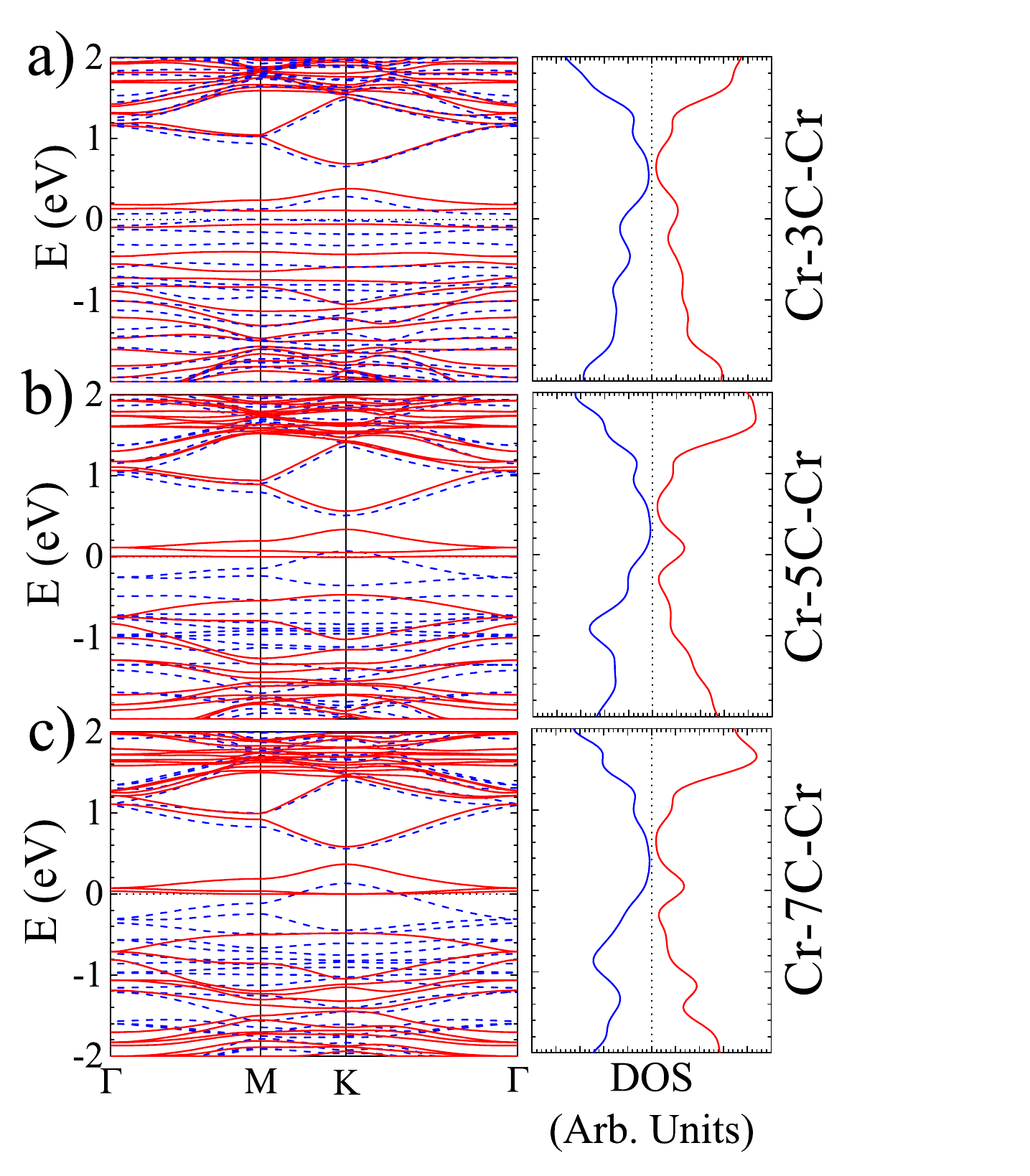}
\caption{The electronic band structure and density of states in chromium-substituted graphene with 3 (a), 5 (b), and 7 (c) separations.}
\label{Cr}
\end{figure}

On the contrary, Figure \ref{Cr} shows the electronic band structure and DOS of the three spatial configurations of the Cr trimers, where the DOS produce a more metallic response. Therefore, the conduction electrons are more active and can provide a stronger exchange energy. It should be noted that, in both cases, the characteristic Dirac cone of graphene is no longer present, which is typically due to the presence of impurity bands and the breaking of inversion symmetry.

In Figure \ref{Cr}(a), we can see the two distinct lowest unoccupied molecular orbital (LUMO) bands and highest occupied molecular orbital (HOMO) band. Once the distance between the magnetic ions increases, the LUMO and HOMO bands start to mix and becomes hard to distinguish (Fig. \ref{Cr}(b-c)). On the other hand, the LUMO and HOMO bands within the Mn case stay distinct for all separations (Fig. \ref{Mn}). One thing to note in the Mn case, as the distance is changed from 3 to 5 carbons, the HOMO band changes from a spin-down to a spin-up and the LUMO band switches from a spin-down to a spin-up.

Within the DOS for 3-separation for Cr (Fig. \ref{Cr}(a)), {one can see a small indication of an impurity resonance peak over the Fermi level, which is a metallic response to the impurity bands in the system}. As the spatial separation is increased (Fig. \ref{Cr}(b-c), the peak becomes more apparent. While in the Mn-substituted graphene (Fig. \ref{Mn}), this peak is suppressed.

\section{Discussion}

Overall, we examine the effect of magnetic trimers on the electronic and magnetic properties of transition-metal substituted graphene. We show that the magnetic interactions between the transition-metal atoms are enhanced by the presence of impurity induced metallic bands. The presence of Mn atom in graphene produces a small energy gap in the DOS. This appears to weaken the exchange interactions between Mn atoms, which can be characterized by a decrease in the induced magnetization of the carbon atoms. With Cr-substituted atoms present, there are distinct impurity bands that produce a metallic response in the DOS. The increase in the DOS seems to enhance the exchange interactions between the Cr atoms, which is characterized by an increase in magnetization in the carbon atoms. 

An examination of the exchange energy as a function of spatial separation for each trimer system reveals that both the Cr and Mn atoms have the same ground states and the both shift from antiferromagnetic to ferromagnetic. This dependence on spatial distance indicates the possibility of an RKKY-like interaction, which is consistent with the difference between the exchange energies of the Mn to Cr systems, where the more metallic Cr-substituted systems produce a larger superexchange. If the exchange is generated through a coupling to the conduction electrons, then it would be expected that more metallic systems will have more substantial coupling.

{The induction of a magnetic moment and RKKY-like spatial dependence is quite similar to that shown with the spin dimer\cite{crook:15}. However, there are distinct differences between the dimer and trimer cases that help reveal the complexity of the interactions in these systems. In the dimer case, the interactions were kept along only the zig-zag direction, which alternates interactions between the two sublattices of the honeycomb lattice. This seems to produce ferromagnetic interactions for magnetic atoms on the same sublattice. In the trimer case, the interactions are along various directions due to the triangular nature of the trimer. If the interactions are completely ferromagnetic on the same lattice site, then the magnetic ground states should be ferromagnetic. However, as shown in Fig. \ref{mag}, when the trimer is close together (as in the 3-carbon separation), this produces a stronger magnetic cluster with antiferromagnetic behavior. This could be due to the more dramatic geometric distortion in the 3-carbon separation case, since the distortion is more localized as the magnetic atoms are separated.}

{Another similarity between the dimer and trimer cases is that the electronic response is different from the Cr to Mn systems. At the Fermi level, the Mn case produces a system with a distinct energy gap, while the Cr case produces an impurity resonance or metallic response. This is likely due to the crystal field effects of each atom in the system. Mn has an extra electron in the $d$-orbitals. Given the magnetic moment, the Cr and Mn atoms are likely in a trigonal pyramidal crystal field and a S = 3/2 state. This would mean that Cr is in 3+ state, while Mn is in a 4+ state. The presence of an extra electron donated by Mn can produce a larger energy gap to electronic system through Coulombic exchange. This different atomic states provide an opportunity to understand the weakening of the exchange interaction in the Mn case and provides a method for experimental verification of these states.}

In conclusion, we show that substitution of magnetic atoms into graphene have the ability to induce magnetism in the carbon atoms. The larger magnetization appears to be connected to the increase in conduction electrons, which in turn produces greater magnetic exchange energy between the trimer atoms. The possibility of an RKKY interaction in these systems means that there is a coupling of electronic and magnetic states, which can lead to electronically controlled magnetic states and the possible realization of spintronic applications in graphene.

\section*{Acknowledgements}

This work was supported by U.S. DOE Basic Energy Sciences E304. G.H., C.B.C., and J.T.H. acknowledge undergraduate support from the Institute for Materials Science and computational support from the Center for Integrated Nanotechnologies at Los Alamos National Laboratory, a user facility. J.T.H would also like to acknowledge travel support from the Kavli Institute for Theoretical Physics at the University of California Santa Barbara. The work at Los Alamos National Laboratory was carried out under the auspice of the U.S. DOE and NNSA under Contract No. DEAC52-06NA25396 and supported by U.S. DOE Basic Energy Sciences Office (J.-X.Z and A.V.B).

\section*{Author's Contributions}

C.B.C., G.H., C.C., J.T.H. contributed to the computational aspects of the project including setup, calculation, and analysis. J.-X.Z and A.V.B provided assistance with analysis and data support. All authors contributed to the writing of the manuscript, although J.T.H and C.B.C provided the main text.

\section*{Funding}

G.H., C.B.C., and J.T.H. received support from the Institute for Materials Science. J.-X.Z and A.V.B are staff researchers at Los Alamos National Laboratory, where their work was carried out under the auspice of the U.S. DOE and NNSA under Contract No. DEAC52-06NA25396 and supported by U.S. DOE Basic Energy Sciences Office.

\end{document}